# The Microlensing Planet Finder:
## Completing the Census of Extrasolar Planets in the Milky Way

D. P. Bennett[a], I. Bond[b], E. Cheng[c], S. Friedman[d], P. Garnavich[a], B. Gaudi[e], R. Gilliland[d], A. Gould[f], M. Greenhouse[g], K. Griest[h], R. Kimble[g], J. Lunine[i], J. Mather[g], D. Minniti[j], M. Niedner[g], B. Paczynski[k], S. Peale[l], B. Rauscher[g], M. Rich[m], K. Sahu[d], D. Tenerelli[n], A. Udalski[o], N. Woolf[i], and P. Yock[p]

[a] University of Notre Dame, Notre Dame, IN, USA
[b] Massey University, Auckland, New Zealand
[c] Conceptual Analytics, LLC, Glen Dale, MD, USA
[d] Space Telescope Science Institute, Baltimore, MD, USA
[e] Harvard-Smithsonian Center for Astrophysics, Cambridge, MA, USA
[f] Ohio State University, Columbus, OH, USA
[g] NASA/Goddard Space Flight Center, Greenbelt, MD, USA
[h] University of California, San Diego, CA, USA
[i] University of Arizona, Tucson, AZ, USA
[j] Universidad Catolica de Chile, Santiago, Chile
[k] Princeton University, Princeton, NJ, USA
[l] University of California, Santa Barbara, CA, USA
[m] University of California, Los Angeles, CA, USA
[n] Lockheed Martin Space Systems Company, Sunnyvale, CA, USA
[o] Warsaw University, Warsaw, Poland
[p] University of Auckland, Auckland, New Zealand

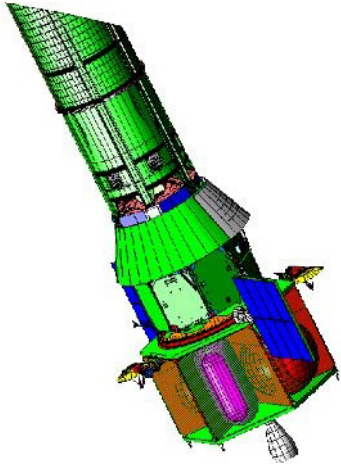

## ABSTRACT

The Microlensing Planet Finder (MPF) is a proposed Discovery mission that will complete the first census of extrasolar planets with sensitivity to planets like those in our own solar system. MPF will employ a 1.1m aperture telescope, which images a 1.3 sq. deg. field-of-view in the near-IR, in order to detect extrasolar planets with the gravitational microlensing effect. MPF's sensitivity extends down to planets of 0.1 Earth masses, and MPF can detect Earth-like planets at all separations from 0.7AU to infinity. MPF's extrasolar planet census will provide critical information needed to understand the formation and frequency of extrasolar planetary systems similar to our own.

The Microlensing Planet Finder's goal is to complete the first census of extrasolar planets in the Milky Way galaxy with sensitivity to planets less massive than the Earth at all separations ≥ 0.7 AU from their host stars. This includes free-floating planets ejected from their host stars. MPF uses the gravitational microlensing technique which is capable of measuring the average number of Earth-like planets per star even if this number is small. NASA's census of Earth-like extrasolar planets will be initiated by the Kepler mission, which has excellent sensitivity to Earth-like extrasolar planets at separations < 0.7 AU as shown in Fig. 1. MPF and Kepler are complementary, with overlap at separations of ~1 AU, in the habitable zone for Solar type stars.

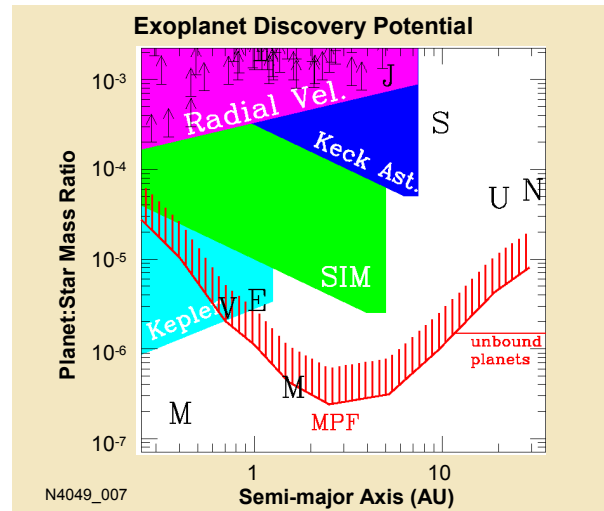

*Fig. 1:* MPF is sensitive to planets above the red curve in the planet:star mass ratio vs. semi-major axis plane. The magenta, blue, green and cyan regions indicates the sensitivities of radial velocity surveys, the Keck interferometer, SIM and Kepler, respectively. The location of our Solar System's planets and many extrasolar planets are indicated.

# 1. MPF Objectives

These are the MPF science objectives:

1. Find 90$f$ planets, where $f$ is the average number of planets per star having the same planet:star mass ratio as the Earth ($q = 3\times10^{-6}$) at orbital separations of 1-2.5 AU. Fig. 2 shows the predicted number of discoveries at 1-2.5 AU.
2. Find 200$g$ planets, where $g$ is the average number of planets per star having the same planet:star mass ratio as an Earth-mass planet orbiting an M-dwarf star ($q = 10^{-5}$) at separations in the range 1-2.5 AU.
3. Find 2800$j$ Jupiter-like planets and 500$s$ Saturn-like planets where $j$ and $s$ are the average number Jupiter-like planets and Saturn-like planets per star. Jupiter-like planets have separations of 5.2 AU and planet:star mass ratios of $10^{-3}$, and Saturn-like planets have separations of 9.5 AU and planet:star mass ratios of $3\times10^{-4}$.
4. For all of the planets discovered by MPF, measure both the planet:star mass ratio and separation to 20% or better. The separation is measured in Einstein ring radius units, and can be converted to physical units with an accuracy of about a factor of 2 with no additional information.
5. If there are $n$ free-floating Earth-mass planets for every star in the Milky Way, MPF will find at least 25$n$ of them. (See Fig. 3.)

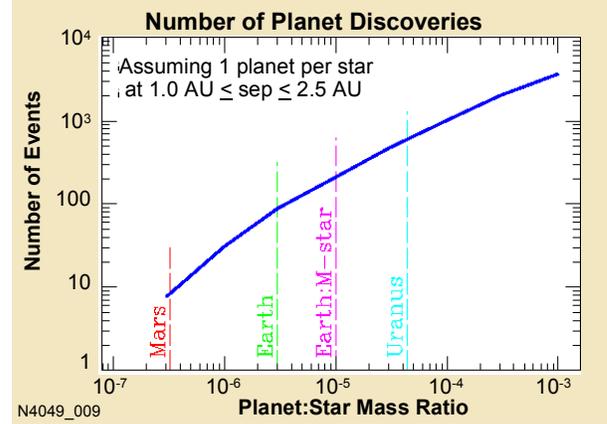

**Fig. 2:** *The expected number of MPF planet discoveries as a function of the planet:star mass ratio if every star has a single planet at a separation of 1.0-2.5 AU.*

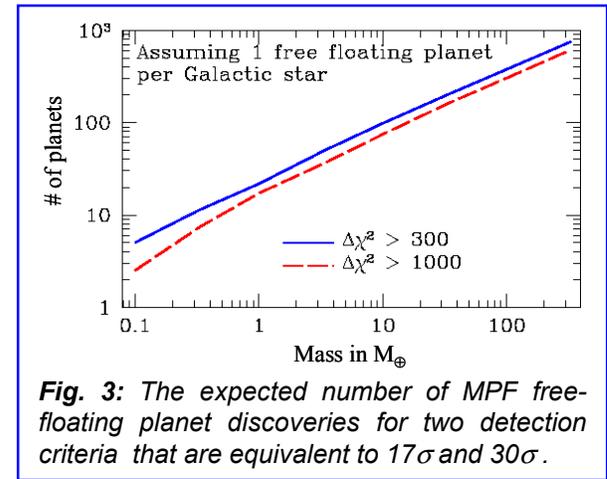

**Fig. 3:** *The expected number of MPF free-floating planet discoveries for two detection criteria that are equivalent to 17$\sigma$ and 30$\sigma$.*

## 1.1. Additional MPF Strengths

In addition to these science objectives which drive MPF's requirements, MPF has the following attributes, most of which follow directly from the requirements and the properties of gravitational microlensing:

- MPF's planetary microlensing signals will have a high signal-to-noise ratio, which will rule out false-positive planet detections without the need for follow-up observations.
- MPF is the only proposed extrasolar planet search program that can find analogs to most of the planets in our own solar system—all but Mercury and Pluto. MPF will discover 66$z$ terrestrial planets, 3300$z$ gas giant planets, and 110$z$ ice giant planets if a fraction $z$ of stars have planets with the same mass ratios and separations as in our own system.
- MPF's sensitivity extends to Mars-like planets at 1/10 of the Earth:Sun mass ratio.
- MPF will measure the frequency of extrasolar planets at the Earth:Sun mass ratio at all separations > 0.7 AU.
- MPF will measure the planet:star mass ratio, which is more fundamental than the planetary area. For example, Saturn's rings have a large area, but a negligible mass.
- The lens stars will be visible for one-third of the planets discovered by MPF including almost all the planets orbiting F, G, or K stars. This will allow an accurate (~20% or better) determination of the lens mass and separation.
- If Earth-like planets are common, MPF will discover its first extrasolar Earths within a few months of launch, since no additional observations are needed to confirm them.
- MPF will measure how the frequency of extrasolar planets varies as a function of Galactic position in the metal-rich portion of our galaxy (the Galactic bulge and inner Galactic disk).
- MPF will discover 150$x$ instances of multiple planets where an average $x$ of lens stars have planetary systems with the same planet:star mass ratios and separations as our own Solar System.
- MPF will detect ~50,000 gas giant planets by transits.

# 2. The Gravitational Microlensing Method

The physical basis of microlensing is the gravitational attraction of light rays by a star or planet. As illustrated in Fig. 4, if a "lens star" passes very close to the line of sight to a more distant source star, the gravitational field of the lens star will deflect the light rays from the source star. The gravitational bending effect of the lens star "splits", distorts, and magnifies the images of the source star. For Galactic microlensing, the image separation is <1 mas, so the observer sees a microlensing event as a transient brightening of the source as the lens star's proper motion moves it across the line of sight.

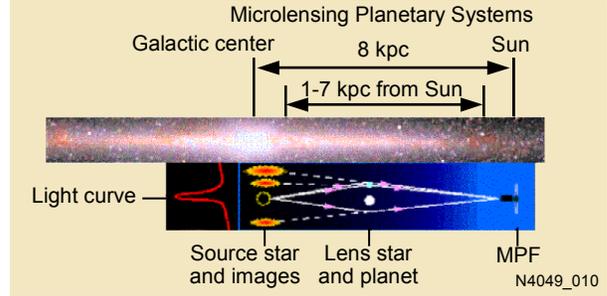

*Fig. 4: The geometry of a microlensing planet search towards the Galactic bulge. Main sequence stars in the bulge are monitored for magnification due to gravitational lensing by foreground stars and planets in the Galactic disk and bulge.*

A gravitational microlensing event is characterized by its Einstein ring radius,

$$R_E = 2.0 \text{ AU} \sqrt{\frac{M_{lens}}{0.5 M_{source}} \frac{D_{lens}(D_{source} - D_{lens})}{D_{source}(1 \text{ kpc})}} ,$$

where $D_{lens}$ and $D_{source}$ are the distances to the lens and source, respectively. This is the radius of the ring image that is seen if the lens and source stars are perfectly aligned. The lensing magnification is determined by the alignment of the lens and source stars measured in units of $R_E$, so even low-mass lenses can give rise to high magnification microlensing events. The duration of a microlensing event is given by the Einstein ring crossing time, typically 1-2 months for stellar lenses, and a few days or less for a planet.

The first gravitational microlensing events were discovered in 1993 by MACHO[1], EROS[2], and OGLE[3], and by now, ~2000 microlensing events have been observed[4,5]. The first microlensing planet discovery was announced in 2004 by MOA and OGLE[6].

## 2.1. The microlensing signals of planets are distinctive and diagnostic.

An extrasolar planet orbiting a lens star becomes detectable when the angular position of the planet comes close to the angular position of one of the images due to the stellar lens, as is shown in Fig. 4. The planet's gravity further deflects the light rays of this image resulting in additional bright images and a deviation of the microlensing light curve from a normal single lens light curve[7]. As shown in Fig. 5, planetary microlensing results in a wide variety of different possible light curve shapes.

Microlensing is most sensitive to planets at a separation of $\sim R_E$ from the lens star. For stellar mass lenses and microlensing events towards the Galactic bulge, $R_E$ is typically 1-5 AU, so this is the region of maximum sensitivity for the microlensing planet search technique[8].

The most important feature of planetary microlensing is that these planetary deviations are large with typical variations of ~10%, even for planets of less than an Earth mass[9]. These signals are the strongest of any proposed Earth-mass planet search technique.

**Microlensing light curves yield unambiguous planet parameters.** For the great majority of events, the basic planet parameters (planet:star mass ratio and planet-star separation) can be "read off" the planetary deviation[8-11].

For a rare (and easily identifiable) subclass of very-high magnification events, there is an irresolvable ambiguity in planet-star separation, although the mass ratio remains well determined[12,13]. This class of events has not been included in our planet discovery estimates. The only other major source of degeneracy in interpretation, which is due to finite-source effects, can be resolved with good quality, continuous light curves[14], such as will be routinely acquired with MPF. The only known effect that can falsely masquerade as a planetary perturbation can also be rejected based on good quality light curves[15]. In brief, virtually all of the planetary discoveries will be unambiguous.

**Planets well below an Earth mass can be detected.** Low-mass planetary microlensing events are more rare, and shorter, than those from more massive planets. Consequently one must observe a much larger sample more frequently to discover these events. The fundamental sensitivity limit occurs when the planet mass becomes too small to magnify more than a small fraction of a star at a time[9]. For main sequence source stars, this limit occurs at about the mass of Mars, so a low-mass planet search program must monitor main sequence source stars.

## 2.2. Detailed simulations of an MPF-like mission have been done.

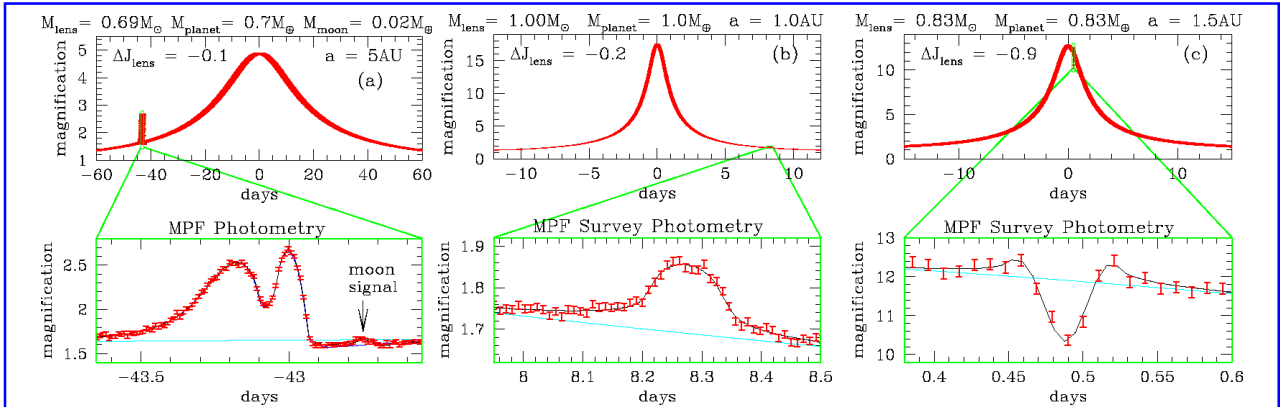

**Fig. 5:** *The light curves of 3 simulated planetary microlensing events with predicted MPF error bars. ΔJ$_{lens}$ is the difference between the lens and source star magnitudes. The lens star is brighter for each of these events. The small deviation at day –42.75 for event (a) is due to a moon of 1.6 lunar masses.*

Detailed simulations of a microlensing planet search satellite mission have been carried out by[16]. These simulations included variations in the assumed capabilities of the satellite that allow us to explore how changes in the mission design will affect the scientific output. Independent simulations by Gaudi (unpublished) have also confirmed the basic results of Bennett and Rhie[16]. These simulation results are the basis for the sensitivity claims made here.

Figure 1 displays MPF's planet detection sensitivity in terms of the planet:star mass ratio, $q$, which is the parameter most directly measured from a planetary microlensing light curve. Since the typical lens star is less massive than the Sun, the typical mass ratio for an Earth mass planet is close to $q = 10^{-5}$. The Earth:Sun mass ratio is $q = 3 \times 10^{-6}$.

Fig. 5 shows light curves typical of what MPF would measure. It is apparent that there are a wide variety of light curve deviations that can be caused by planets. The relatively high S/N of the microlensing planet detections allows us to distinguish planetary microlensing from other types of light curve variations, and to accurately determine the planetary parameters.

### 2.3. Many planetary host stars are visible, including Solar type.

The planets detected by the MPF survey orbit the lens stars in the foreground of the Galactic bulge source stars. The mass distribution of the lens stars from our MPF simulations is shown in Fig. 6. This distribution is somewhat flatter than the stellar mass function because we have (conservatively) assumed that the planetary mass distribution is proportional to the stellar mass distribution and more massive planets are more likely to be detected.

Although microlensing does not require the detection of any light from the lens stars, a significant fraction of the microlensing events seen by MPF will have lens stars that are bright enough to be detected. Our simulations indicate that for ~17% of the detected planets, the planetary host (lens) star is brighter than the source star, and for another ~23% the lens star's brightness is within 2.5 I-band magnitudes of the source star. A few of these stars are blended with the images of other brighter stars, and if we ignore those stars, we find that 33% of the lens stars should be directly detectable. The detectable planetary host stars are depicted in red in Fig. 6 and they comprise virtually all of the F and G star lenses, most of the K star lenses, and a few of the nearby M star lenses[17,18].

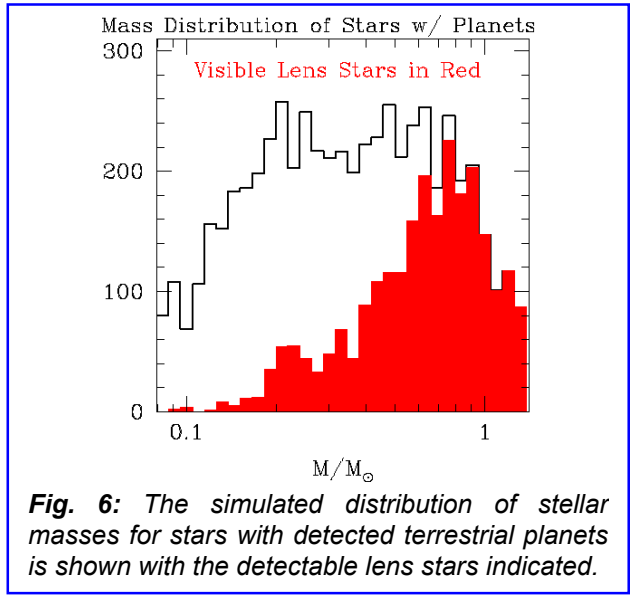

**Fig. 6:** *The simulated distribution of stellar masses for stars with detected terrestrial planets is shown with the detectable lens stars indicated.*

For the one-third of lens stars that are visible, it will be possible to estimate the lens star mass and distance from its dereddened color and magnitude (and the fact that essentially all visible lenses are on the main sequence). This has already been demonstrated with HST images of one microlensing event[19-21]. The planetary mass can then be derived from the lens star mass and the planet:star mass ratio (which is routinely determined from the light curve). Similarly, the physical planet-star projected separation can be found from $R_\perp = u_p R_E$, where the Einstein radius $R_E$ is derived from the photometric lens and source distances and the lens mass, and $u_p$ is the normalized planet-star separation (which is also determined from the light curve). In most cases, the extinction to the source and lens can be adequately estimated using the standard microlensing technique summarized by Yoo et al.[22]. These estimates can be improved by ground-based high-resolution observations using adaptive optics (AO) on VLT, Gemini, Keck, and LBT (to the last of which we have substantial dedicated access).

While this approach will yield mass estimates accurate to 10-20%, the resulting projected separations may not be accurate for a subset of planets lying in the Galactic bulge. However, the uncertainty in these cases can be overcome by measuring the *angular* Einstein radius $\theta_E$ (and so $R_\perp = u_p D_{lens} \theta_E$) which can be done in either of two ways. First, the small lens-source relative proper motion $\mu_{rel} \sim 7$ mas/yr will cause the lens and source to separate by ~20 mas during the mission. If this can be measured from the resulting distortion of the PSF (for lens and source stars of similar brightness), it will yield a measurement of $\mu_{rel}$ (and so $\theta_E = \mu_{rel} t_E$, where $t_E$ is the Einstein crossing time, which is measured during the event). If the PSF distortion is too subtle to measure, the source-lens separation will be measurable from the ground with AO about 10 years after the event. Second, whenever the source passes close to or over a "caustic" generated by the lens, it is possible to measure $\theta_E$ (see Yoo et. al[22]). Such caustic crossings occur frequently for planetary events simply because the source must pass close to the planetary caustic for the planet to be detectable at all. The planetary microlensing event detected from the ground had such a caustic crossing[6].

Finally, it is possible to determine the masses and separations of many lens stars and planets (including some that are not visible) using the microlensing parallax effect[23,24].

### 2.4. Microlensing can detect free-floating planets.

A unique capability of microlensing is its ability to detect free-floating planets[25,26]. These can be detected down to a Mars mass as very short timescale microlensing events. Such planets are expected to be very common because the late stages of planetary system formation involve planetary scattering events in which a close approach between two planets results in one of the planets being ejected from the planetary system. It is also thought that giant planets like Jupiter will routinely eject a large number of planets in the terrestrial mass range[27] during the early phases of planetary accretion.

An important consequence of the sensitivity to free floating planets is that, should Earth-like planets prove to be rare, the free-floating planet survey might provide an explanation. If each star ejected one Earth mass planet, MPF would detect about 25 such objects (See Fig. 3).

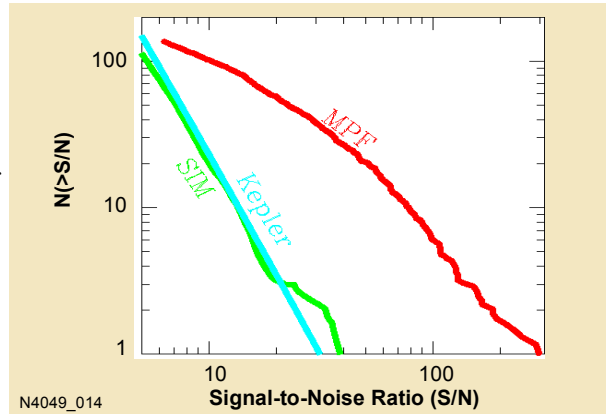

**Fig. 7:** A comparison of the signal-to-noise ratio distribution for missions able to detect Earth-like planets, which are assumed to have the Earth:Sun mass ratio.

### 2.5. Earth-like planets are discovered by microlensing with a high S/N.

A high signal-to-noise ratio (S/N) is valuable for ruling out false positive detections. Models of the discovery can be compared directly to models of the false positive signals, but if the S/N is too low, then only the null hypothesis can be ruled out. Even if no such models are known, a high S/N ratio allows a confrontation with the discovery model. Discoveries made with low S/N require an additional effort to rule out false positives.

Fig. 7 shows the S/N distributions for Earth-like planet discoveries made by MPF, SIM, and Kepler[28]. In all cases, planets with the Earth:Sun mass ratio of $3\times10^{-6}$ are assumed. For Kepler, each planet is located at the center of its habitable zone; for MPF, it is located at 1-2.5 AU; and for SIM, it has a 3-year period. For SIM and Kepler, the number of detections with a signal-to-noise ratio exceeding some threshold scales as $\sim(S/N)^{-3}$. Such a scaling is expected for a mission where the number of target stars scales as their distance cubed.

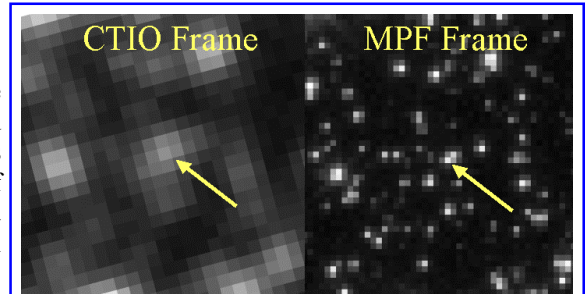

**Fig. 8:** A comparison between an image of the same star field in the Galactic bulge from CTIO in 1" seeing and a simulated MPF frame (based on an HST image). The indicated star is a microlensed main sequence source star.

For MPF, all the source stars are at a similar distance, so the S/N distribution depends on the bulge luminosity function and on the intrinsic properties of the microlensing events. Our rate scales as $\sim (S/N)^{-1.1}$ for $12.5 < S/N < 60$, so a large fraction of MPF's planet discoveries will have a S/N much greater than our threshold of 12.5. This means that we will generally have enough S/N in the data to rule out any potential false positive detections.

**The microlensing technique is robust to descopes.** The shallow slope of MPF's S/N curve in Fig. 7 offers an important advantage. If we are forced to loosen a requirement, it may have only a small effect on the number of planets discovered. For example, a drastic 20% reduction in the detector QE would result in only 11% fewer MPF planet discoveries.

### 2.6. Ground Based Microlensing Surveys cannot do MPF's mission.

Gravitational microlensing has developed as a ground based observational technique over the past decade. The EROS, MACHO, and OGLE microlensing survey teams have observed ~2000 microlensing events caused by ordinary stars along the line of sight to the Galactic bulge[4,5], and one case of microlensing by a ~1.5 Jupiter mass planet has been found[6]. A previous reported planet detection[29] was misinterpreted[30] due to data taken in poor seeing.

There have been several suggestions that ambitious ground-based microlensing surveys might be a cost effective method to find Earth-mass planets from the ground[31,32], but these failed to appreciate the extreme crowding of main sequence source stars in the Galactic bulge (see Fig. 8). (Only main sequence stars have an angular size that is small enough not to wash out the photometric signal from Earth-like planets[9].) Detailed calculations that have taken this effect into account have shown that ground-based surveys cannot detect a significant number of terrestrial planets[33,34]. Bennett[34] simulated a network of 2m wide FOV telescopes, which can image 7 deg$^2$ in a single image. He found that the number of Earth-mass ratio planets discovered at orbital radii of 0.5-1.5 AU is a factor of ~100 less than the number expected from a space-based survey, and there is little improvement with larger ground-based telescopes due to systematic photometry errors. The sensitivity is compared to that of MPF in Fig. 9. The problem with ground-based surveys is due in part to the need for continuous observations on a ~24-hour time scale. This implies observations from multiple ground-based sites, which can often be clouded out or have less than excellent seeing. This approach often leads to detectable (raw) planet signals that do not yield unique planetary parameters, as shown in Fig. 10.

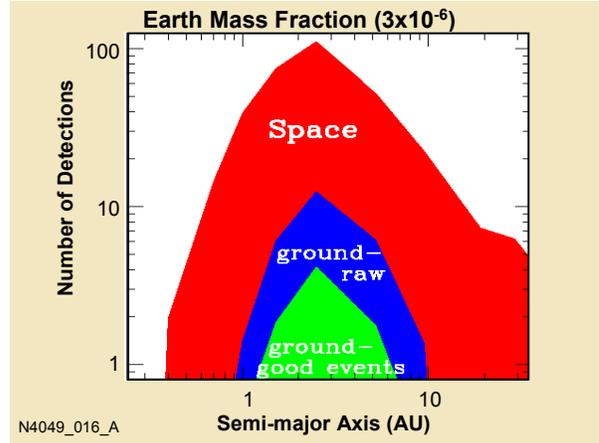

**Fig. 9:** The number of expected Earth-like planet detection is plotted as a function of orbital semi-major axis for MPF (space) and a 3-site wide FOV ground based survey. The green region indicates the events detected and well characterized from the ground, while the blue region indicates detectable events poor light curve coverage.

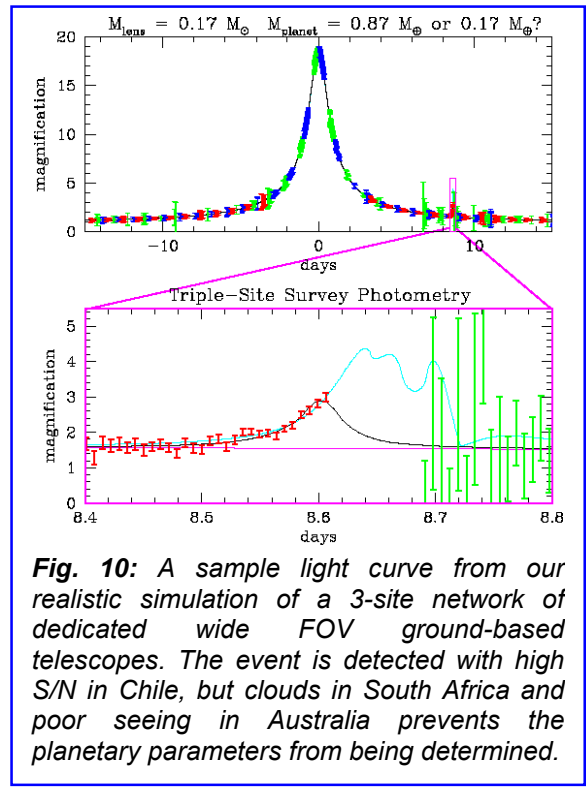

**Fig. 10:** A sample light curve from our realistic simulation of a 3-site network of dedicated wide FOV ground-based telescopes. The event is detected with high S/N in Chile, but clouds in South Africa and poor seeing in Australia prevents the planetary parameters from being determined.

### 3. Overview of the MPF Mission and Instrument

Key requirements of the MPF mission are summarized in table 1. MPF will continuously observe two 1.3 deg$^2$. fields in the central Galactic bulge with a sampling interval of 15 minutes for each field. MPF requires an orbit allowing nearly continuous monitoring of its Galactic bulge target fields, interrupted only as necessary, such as a 3-month period every year to avoid the Sun. The optimal orbit is a geosynchronous orbit inclined by 28.7° to the Equator and 52° to the Ecliptic, as shown in Fig. 11. This allows a continuous view of the Galactic bulge for all but 1-2 days per month plus continuous data downlink to a dedicated MPF ground station in White Sands, NM. The orientation of the MPF telescope is kept fixed for the first 4.5 months of the 9-month MPF observing season, and then, the spacecraft is rolled by 180° about the telescope bore-sight. This prevents sunlight from falling on the "dark site" of the telescope where the thermal radiators are mounted, since the MPF field is only ~5° from the Ecliptic plane.

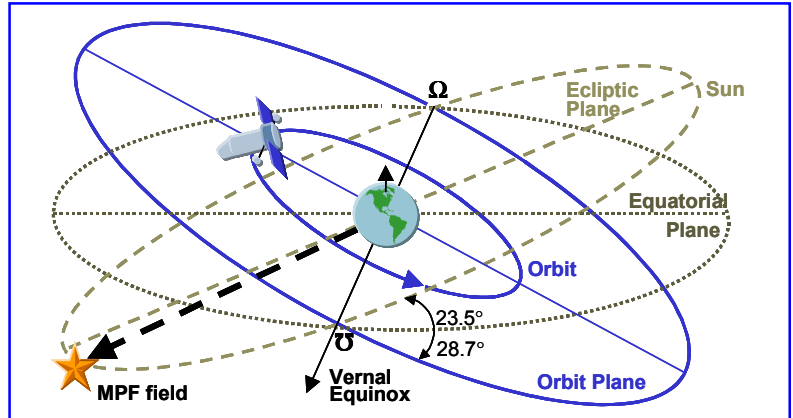

*Fig. 11: The MPF orbit is inclined by 28.7° to the Equator and ~52° to the Ecliptic plane. This allows for a continuous data downlink, a >40° angle between the Earth's limb and the line-of-sight to the MPF target field and a low energy launch.*

The orientation of the two MPF fields on the Galactic bulge is shown in Fig. 12. The MPF focal plane consists of two 5×7 banks of detectors separated by a distance equal to the width of a single detector bank. This allows the two MPF pointings to cover a continuous region of 0.96° × 2.73°. The orientation of the MPF focal plane allows two MPF pointings to be aligned with the Galactic plane as indicated in Fig. 12. This allows the MPF fields to be selected to maximize the microlensing event rate and extrasolar planet detection sensitivity.

Each MPF pointing includes 6 × 62 s exposures using MPF's array of 70 2048$^2$ detectors. 80% of the detectors are HgCdTe IR detectors with a 1.7 μm long wave cutoff (such as the IR detector used for HST/WFC3), and 20% are Si PIN optical detectors. The QE curves shown in Fig. 13 indicate that the HgCdTe IR arrays are, by far, the detectors most sensitive to MPF's target stars, which are reddened main sequence stars in the Galactic bulge. The Si PIN detectors are included to provide some color information. The internal boundaries of MPF's two passbands are set by the detector QE curves, while the outer limits

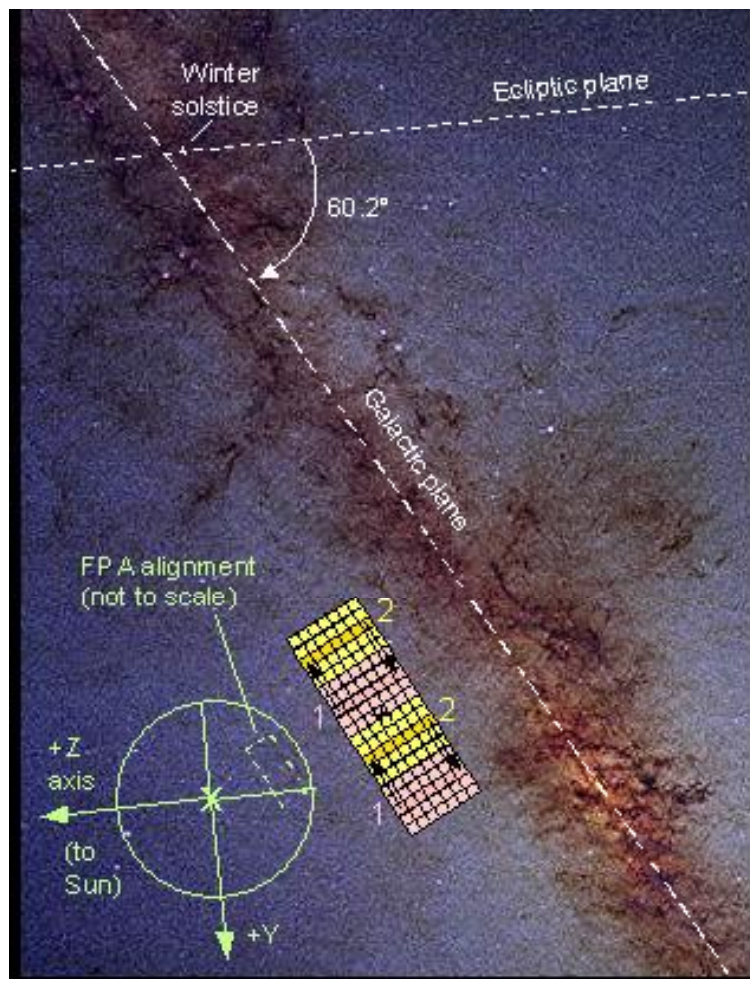

*Fig. 12: The orientation of the MPF fields on the 2-MASS image of the Galactic bulge.*

| Property | Value | Units |
|---|---|---|
| Launch Vehicle | 7920-10L | Delta II |
| Orbit | Inclined GEO 28.7 | degrees |
| Mission Lifetime | 3.7 | years |
| Telescope Aperture | 1.1 | meters (diameter) |
| Field of View | 1.25 | square degrees |
| Spatial Resolution | 0.240 | arcsec/pixel |
| Pointing Stability | 0.024 | arcsec |
| Focal Plane Format | 290 | Megapixels |
| Spectral Range | 600 – 1600 | nm in 2 bands |
| Quantum Efficiency | > 75% <br> > 65% | 600-950 nm <br> 1000-1600 nm |
| Dark Current | < 1 | e-/pixel/sec |
| Readout Noise | < 20 | e-/read |
| Photometric Accuracy | 1 or better | % at I=21.5. |
| Data Rate | 42 | Megabits/sec |

*Table 1: Key Mission Requirements*

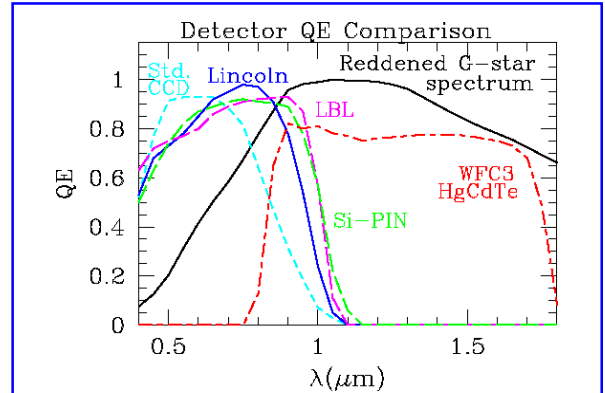

*Fig. 13: The spectrum of a typical MPF target star is compared to the quantum efficiency (QE) curves for different detector types. MPF's WFPC3 style HgCdTe detectors are optimal.*

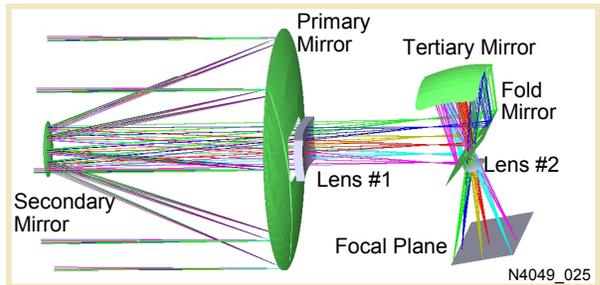

*Fig. 14: MPF Telescope Configuration.*

are set to be 600-1600 nm with a filter.

The 290 million MPF pixels result in a peak raw data rate of 76 Mbits/s before compression. The MPF detectors allow for non-destructive readouts which require no dead time between exposures at the same pointing. They also require no shutter and represent a substantial improvement in sensitivity over the previously proposed design for the GEST mission[35], which had very similar science goals.

Each MPF pixel subtends 0.24 arcsec, which corresponds to the diffraction limit FWHM at a wavelength of ~1.3 μm and is sufficient to resolve most of the bulge main sequence stars. At least once a week, images are taken with offset pointings to image each star in the fields with the different passbands defined by the two detector types. Each MPF pointing is slightly offset from the previous pointing to the same field in a dither pattern designed to produce photometry with accuracy better[36,37] than 1%.

MPF employs a 1.1m aperture three mirror anastigmatic telescope (TMA), as shown in Fig. 14. This design delivers diffraction limited optical performance over MPF's wide wavelength range. Two lenses have been added to the traditional TMA design, primarily for the purpose of radiation shielding.

The radiation environment in a geosynchronous orbit presents a challenge for MPF. This orbit places MPF in the outer part of the electron Van Allen belt which exposes the spacecraft to a high dose of low energy electron radiation. These electrons present little radiation damage risk, and they can be stopped with a modest amount of radiation shielding. However, the primary electrons can produce secondary x-rays when they interact with the shielding material, and both the electrons and x-rays can interfere with MPF's science by producing "glitches" in the MPF images when they interact with the MPF detector pixels. This problem can be solved by completely enclosing the MPF detectors in a radiation shield as shown in the outline in Fig. 15. This tee-pee shaped radiation shield is closed at the top by lens #2 as indicated in Fig. 14. Detailed calculations indicate that the detector glitch rate can be reduced to a level that is almost as small as the rate expected from cosmic ray protons with this lens plus a radiation shield composed of 2 g/cm$^2$ of a lead-aluminum "sandwich". This tee-pee shield also serves as part of the thermal control system that allows the detectors to be passively cooled to ~140K.

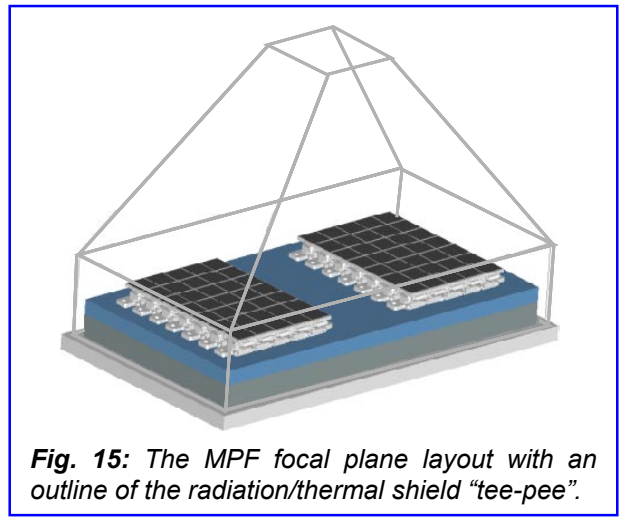

*Fig. 15: The MPF focal plane layout with an outline of the radiation/thermal shield "tee-pee".*

Once the data are transmitted to the ground, they are processed into photometry by a method used for HST data with similar image sampling[37]. A set of light curves of variable objects is produced, which include both the microlensing and transit events. The microlensing events are identified and characterized to reveal the properties of the detected planets using methods developed by ground-based microlensing programs[4-6,38-42]. The analysis of the planetary microlensing light curves leads directly to all of the science outputs described in Sec. 1.

## 4. Follow-up Observations of Microlensing Planet Discoveries

About one-third of the host stars of MPF-detected planets will be at least 10% as bright as the source and therefore visible. For these stars, important auxiliary information can be obtained from the ground. Most critical is IR photometry at 2 epochs, one during the event, while the source star is still magnified, and one after the event, to estimate the lens mass and distance. Whenever possible, it will be important to obtain the radial velocity (RV) and proper motion (to derive the Galactic orbit) and the metallicity. These will permit measurement of planet frequency as simultaneous functions of Galactic position and metal abundance. We expect the proper motions can be derived from the MPF data alone, but if not, they can be obtained from late-time AO IR imaging about 10 years after the lensing event.

For the one-sixth of lenses that are brighter than their source, RV and metallicity can be efficiently obtained from multiplexed AO IR spectroscopy after the event (using e.g., SPIFFI/VLT, LUCIFER/LBT, OSIRIS/Keck, or GNIRS/Gemini). Several MPF CoI's have substantial dedicated time on LBT (Bennett, Gould) and have routine access to Keck and all Chilean telescopes.

## 5. Pre-Launch Observations

Upon approval by NASA, we will characterize the fields using ground-based observations. Candidate MPF fields will be monitored by the OGLE-III and MOA-II microlensing surveys to allow us to select the fields with the highest rate of observable microlensing events. The candidate fields will also be imaged at the highest possible resolution with one of the large telescopes located in Chile, in the optical and near-IR. These images will provide an accurate map of the stellar populations, deep luminosity function, reddening, etc., with multicolor characterization of each potential microlensing source star. They will also provide accurate multiband photometric calibrations of thousands of relatively uncrowded standard stars in our fields, allowing accurate calibration of MPF's very wide passbands in space with the data from the primary science fields. These images would also serve as finding charts for the follow-up spectroscopic observation of microlensing events.

## 6. Discussion and Summary

Mankind's interest in extrasolar planets is driven largely by our interest in extra-terrestrial life and by the fact that the only life we know about has originated on our own planet. The ultimate goal of understanding extra-terrestrial life means that Earth-like planets are of particular interest. However, our ignorance of the origin of life is nearly complete, and our understanding of the planet formation process is quite primitive. It is fair to say that we have only a very crude idea about what is needed for a planet to be habitable[41-44] and to support the development of extra-terrestrial life[45-49]. Thus, it is sensible to first study the basic properties of planetary systems and to try to develop an understanding of how planetary systems form before focusing too closely on Earth-like planets. For these reasons, the latest Decadal Survey[50] has recognized that the next step, after the discovery of the first extrasolar planet orbiting main sequence stars[51], is to carry out a census of extrasolar planets in order to determine the basic properties of planetary systems. Discoveries of extrasolar giant planets made to date[52-54] suggest that many extrasolar planetary systems may be incompatible with the formation or survival of Earth-like planets[55-57]. Thus, it is important to carry out an extrasolar planet census that will be able to measure the frequency of Earth-like planets, even if this number is relatively small, say 10%.

MPF is the ideal mission to complete such a census. The MPF mission is optimized to:
- Provide the first census of planets like those in our own solar system, with sensitivity to analogs of all solar system planets except for Mercury and Pluto.
- Discover 66 terrestrial planets, 3300 gas giants, and 110 ice giant planets (if our solar System is typical).
- Discover Earth-like planets at 1-2.5 AU from their stars even if only a few per cent of stars have such planets.

- Discover Earth-like planets within a few months of launch.
- Discover free-floating planets, not gravitationally bound to any star.
- Provide direct measurement of star:planet mass ratios, plus planet masses and separations for planets orbiting solar-type stars.

The only region where MPF has poor sensitivity to extrasolar planets is at separations of < 0.7 AU, but this is the region where Kepler is most sensitive[58]. Thus, the combination of Kepler and MPF will provide measurements of the frequency of Earth-like and larger extrasolar planets at all star:planet separations and will complete the first census of extrasolar planets like those in our own Solar System.

## ACKNOWLEDGMENTS

We would like to thank Ken Ford, John Azzolini (Goddard Space Flight Center), Kin Chan, Bob Woodruff, Andy Klavins, Tom Sherrill (Lockheed Martin), Tom Chuh, Dave Gulbransen (Rockwell Scientific), Jeff Wynn (Kodak), Paul Nikulla (Swales), and Gerry Luppino (GL Scientific) for help with the development of the MPF mission concept. Theoretical calculations in support of the MPF mission were supported, in part, by NASA Origins Grant NAF5-13042.